# A Comparative Analysis of Machine Learning Techniques for IoT Intrusion Detection

João Vitorino[0000-0002-4968-3653], Rui Andrade[0000-0003-2356-3706], Isabel Praça[0000-0002-2519-9859], Orlando Sousa[0000-0003-0779-3480] and Eva Maia[0000-0002-8075-531X]

Research Group on Intelligent Engineering and Computing for Advanced Innovation and Development (GECAD), School of Engineering, Polytechnic of Porto (ISEP/IPP), 4249-015 Porto, Portugal
{jpmvo,rfaar,icp,oms,egm}@isep.ipp.pt

**Abstract.** The digital transformation faces tremendous security challenges. In particular, the growing number of cyber-attacks targeting Internet of Things (IoT) systems restates the need for a reliable detection of malicious network activity. This paper presents a comparative analysis of supervised, unsupervised and reinforcement learning techniques on nine malware captures of the IoT-23 dataset, considering both binary and multi-class classification scenarios. The developed models consisted of Support Vector Machine (SVM), Extreme Gradient Boosting (XGBoost), Light Gradient Boosting Machine (LightGBM), Isolation Forest (iForest), Local Outlier Factor (LOF) and a Deep Reinforcement Learning (DRL) model based on a Double Deep Q-Network (DDQN), adapted to the intrusion detection context. The most reliable performance was achieved by LightGBM. Nonetheless, iForest displayed good anomaly detection results and the DRL model demonstrated the possible benefits of employing this methodology to continuously improve the detection. Overall, the obtained results indicate that the analyzed techniques are well suited for IoT intrusion detection.

**Keywords:** internet of things, intrusion detection, supervised learning, unsupervised learning, reinforcement learning

## 1 Introduction

The digital transformation is associated with the Internet of Things (IoT) concept, which describes decentralized and heterogeneous systems of interconnected devices. This field converges wireless sensor networks, real-time computing, embedded systems and actuation technologies [1]. Industrial IoT (IIoT) is a subfield of IoT focused on industrial assets and the automation of manufacturing processes. Due to the integration of physical and business processes, as well as control and information systems, IIoT is bridging the gap between Operational Technology and Information Technology [2].

However, the convergence of previously isolated systems and technologies faces tremendous security challenges. IoT devices commonly have software and communication protocol vulnerabilities, in addition to weak physical security and resource constraints [3], [4]. Consequently, malware attacks pose a major threat to IoT systems. A self-propagating malware, such as Mirai, can compromise a large number of susceptible devices and establish a botnet to launch several cyber-attacks [5]. The cyber-attacks targeting IoT systems can be divided into two categories: passive and active. Passive attacks do not impact the operation of the system, mainly consisting of eavesdropping and traffic analysis. On the other hand, active attacks can range from probing and man-in-the-middle to brute-force and Denial-of-Service (DoS) [6], [7].

Due to the exposure of IoT to malicious activity, a reliable intrusion detection is indispensable. An Intrusion Detection System (IDS) dynamically monitors an environment with the purpose of identifying suspicious activity, so that possible threats can be mitigated [8]. The application of machine learning techniques to an IDS is a promising strategy to tackle the growing number and increasing complexity of cyber-attacks.

The developed work addressed nine malware captures of the IoT-23 dataset in both binary and multi-class classification scenarios. Three distinct types of techniques were analyzed and compared: supervised, unsupervised and reinforcement learning. The developed models consisted of three supervised models, Support Vector Machine (SVM), Extreme Gradient Boosting (XGBoost) and Light Gradient Boosting Machine (LightGBM), two unsupervised models, Isolation Forest (iForest) and Local Outlier Factor (LOF), and a Deep Reinforcement Learning (DRL) model based on a Double Deep Q-Network (DDQN), adapted to the intrusion detection context.

This paper is organized into multiple sections. Section 2 provides a survey of previous work on machine learning techniques for intrusion detection. Section 3 describes the utilized dataset and models, including the data preprocessing steps and evaluation metrics. Section 4 presents an analysis of the results obtained in each scenario. Finally, Section 5 addresses the main conclusions and future research topics.

## 2    Related Work

In recent years, IoT intrusion detection has drawn attention from a research perspective. As both cyber-attacks and the techniques used to detect them evolve, an increasing number of research topics come to light. Therefore, it is essential to understand the results and conclusions of previous work.

Chaabouni et al. [8] provided a comprehensive survey of research published up to the year of 2018. The authors reviewed previous studies aimed at IoT, highlighting the advantages and limitations of the developed machine learning models.

More recently, Zolanvari et al. [9] utilized a testbed mimicking an industrial plant to train several models for anomaly detection. The best overall performance was achieved by Random Forest, which obtained a True Positive Rate (TPR) of 97.44%. However, only SVM reached a False Positive Rate (FPR) of 0.00, representing no false alarms.

Jan et al. [10] proposed the use of SVM to detect attacks that influence IoT network traffic intensity, which is common in DoS. The performance of different SVM kernels was analyzed on simulated datasets with only three features: the minimum, maximum and median values of the packet arrival rate. Even though the Linear kernel reached 98.03% accuracy with a small training time, this approach lacks the ability to detect attacks that do not increase neither decrease traffic intensity.

Bakhtiar et al. [11] employed the lightweight C4.5 algorithm to search for DoS attacks by directly analyzing the packets captured in a device and creating a decision tree. Despite achieving an accuracy of 100%, the average time required to process each one was 0.0351 seconds on their testbed. Consequently, only 18.15% of the transmitted packets were analyzed, which revealed the drawback of a packet-based approach.

Verma and Ranga [12] addressed classifier ensembles, comparing several models on the CIDDS-001, UNSW-NB15 and NSL-KDD datasets. 10-fold cross-validation was performed and the highest average accuracy, 96.74%, was obtained by the Classification And Regression Trees algorithm, which creates a decision tree. However, XGBoost reached the very close value of 96.73% and obtained the best average TPR, 97.31%.

Yao et al. [13] proposed the use of LightGBM to perform a lightweight analysis in IoT devices, followed by more resource-intensive models in other devices. The authors noted that since LightGBM is embedded with feature selection, the bandwidth required to transmit the data is reduced. On their dataset, LightGBM achieved an accuracy of 93.2% and an F1-score of 95.6% for a flow-based approach.

Eskandari et al. [14] used unsupervised models to perform anomaly detection by building a baseline of benign flows. LOF and iForest were compared in their testbed with probing, brute-force and DoS attacks. Their macro-averaged F1-scores were 78.4% and 92.5%, respectively, which indicates the suitability of the latter for the detection of unknown attacks when trained with normal network traffic only.

The key drawback of both supervised and unsupervised techniques is that if the cyber-attacks are modified or the network topology is updated, which includes the addition of a new device, the models must be retrained to take into consideration the new traffic patterns. To tackle this challenge, reinforcement learning can be adapted to the intrusion detection context.

Gu et al. [15] proposed an entropy-based approach to continuously optimize a threshold for anomaly detection. An agent interacted with the network environment, receiving TPR and FPR as the rewards for each selected threshold. It employed Q-Learning, which is an off-policy learner because it is improved regardless of the agent's actions.

Despite not being aimed at IoT, Lopez-Martin et al. [16] analyzed the performance of several techniques that combine reinforcement learning with deep learning to create DRL models with improved stability. The utilized agents directly predicted the class of the network flows received from the environment. Regarding the reward function, the authors noted that a simple 1/0 reward for correct/incorrect predictions led to a better performance. The best results were achieved by a DDQN, with F1-scores of 91.20% and 93.94% on the NSL-KDD and AWID datasets, respectively.

To the best of our knowledge, no previous work has comparatively analyzed supervised, unsupervised and reinforcement learning techniques on the IoT-23 dataset.

## 3   Methods

This section describes the utilized dataset and models, as well as the employed data preprocessing steps and the considered evaluation metrics. The work was carried out on a machine with 16GB of RAM, an 8-core CPU and a 6GB GPU. The implementation relied on the Python programming language and the following libraries: *Numpy* and *Pandas* for general data manipulation, *Scikit-learn* for the implementation of SVM, iForest and LOF, *Xgboost* for the implementation of XGBoost, *Lightgbm* for the implementation of LightGBM and *Tensorflow* for the implementation of the DRL model.

### 3.1    Dataset

The IoT-23 dataset [17] was created by the Stratosphere Research Laboratory and is publicly available. It consists of twenty-three labeled captures of malicious and benign network flows, caused by malware attacks targeting IoT devices between 2018 and 2019. This is an extremely valuable dataset because it manifests real IoT network traffic patterns and provides a large quantity of labeled malicious flows.

From the twenty-three captures, six were selected due to their distinct characteristics. Since Capture-1-1 displayed a large number of recorded flows and the best balance between malicious and benign labels, it was renamed as 1-1-full and three smaller balanced subsets were established: 1-1-large, 1-1-medium and 1-1-small. Table 1 provides an overview of the malware type and class proportions of the utilized datasets. The labels PartOfAHorizontalPortScan and C&C-FileDownload were shortened to POAHPS and C&C-FD, respectively.

**Table 1.** Main characteristics of the utilized datasets.

| Dataset | Malware Type | Total Samples | Malicious Class Samples | Malicious Class Label |
|---|---|---|---|---|
| 1-1-full | Hide and Seek | 1,008,749 | 539,465 | POAHPS |
|  |  |  | 8 | C&C |
| 1-1-large | Hide and Seek | 400,000 | 199,996 | POAHPS |
|  |  |  | 4 | C&C |
| 1-1-medium | Hide and Seek | 200,000 | 99,999 | POAHPS |
|  |  |  | 1 | C&C |
| 1-1-small | Hide and Seek | 20,000 | 10,000 | POAHPS |
| 20-1 | Torii | 3,210 | 16 | C&C-Torii |
| 21-1 | Torii | 3,287 | 14 | C&C-Torii |
| 34-1 | Mirai | 23,146 | 14,394 | DDoS |
|  |  |  | 6,706 | C&C |
|  |  |  | 122 | POAHPS |
| 42-1 | Trojan | 4,427 | 3 | FileDownload |
|  |  |  | 3 | C&C-FD |
| 44-1 | Mirai | 238 | 14 | C&C |
|  |  |  | 11 | C&C-FD |
|  |  |  | 1 | DDoS |

### 3.2   Data Preprocessing

Besides the creation of the three additional subsets, a preprocessing stage was required before the data was usable (see Fig. 1). This stage was applied to all nine datasets, taking into consideration their distinct characteristics.

A pertinent aspect is that if a class only contains a single sample, it cannot be simultaneously used to train and evaluate a model. Therefore, that individual sample must be discarded. This is the case of the 1-1-medium and 44-1 datasets, when used for multi-class classification. Regarding 1-1-medium, it becomes only suitable for binary classification because only the POAHPS malicious class remains. Consequently, only 1-1-full, 1-1-large, 34-1, 42-1 and 44-1 were utilized in the multi-class scenario.

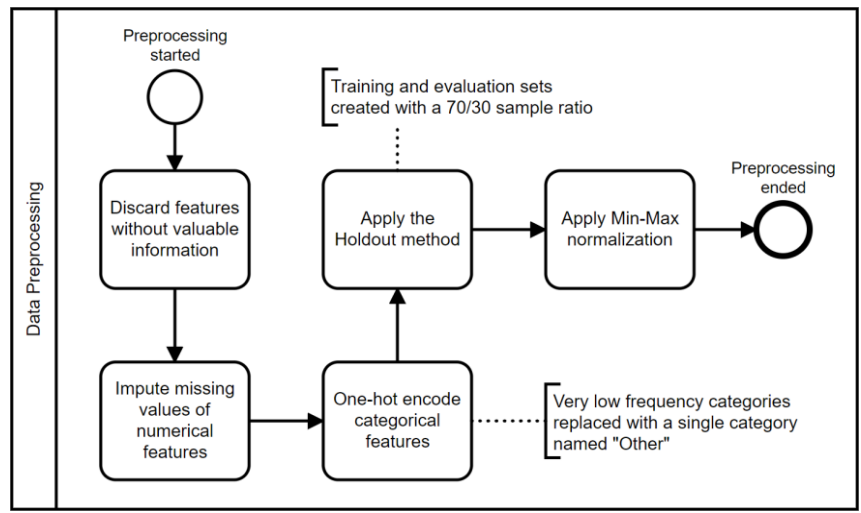

**Fig. 1.** Overview of data preprocessing stage (Business Process Model and Notation).

### 3.3   Evaluation Metrics

The performance of a model can be evaluated using the values reported by the confusion matrix. It reports the number of True Positives (TP), True Negatives (TN), False Positives (FP) and False Negatives (FN) regarding the predicted classes. Using binary classification as an example, the considered metrics and their interpretation are described below [18], [19].

Accuracy measures the proportion of correctly classified network traffic. However, a high value can be achieved even when a minority class is disregarded. For instance, a high accuracy can be reached in datasets unbalanced towards benign traffic without any malicious activity being detected.

Precision measures the proportion of predicted attacks that were actual attacks, which indicates the relevance of a model's predictions. On the other hand, Recall, which corresponds to TPR, measures the proportion of actual attacks that were correctly predicted, reflecting a model's ability to identify malicious activity. FPR is another valuable metric because it accounts for false alarms, which must be avoided. It measures the proportion of benign traffic that was incorrectly predicted to be an attack, leading to unnecessary mitigation efforts.

Overall, the most trustworthy metric is the F1-score, also referred to as F-measure. It calculates the harmonic mean of Precision and Recall, considering both FP and FN. Therefore, a high F1-score indicates that malicious activity is being correctly identified and there are low false alarms.

These metrics, except for Accuracy, can be macro-averaged to treat all classes equally. Since the minority classes are given the same relevance as the overrepresented, macro-averaging is well suited for unbalanced datasets.

### 3.4  Supervised Learning Models

Due to the promising results obtained in the surveyed work, three supervised techniques were selected to be evaluated on the IoT-23 dataset. The configurations of the developed models resulted from a grid search of possible hyperparameter combinations for both binary and multi-class classification scenarios.

To obtain the optimal configuration for each dataset and scenario, a 5-fold cross-validation was performed. Therefore, a model was trained with 4/5 of a training set and validated with the remaining 1/5 in each iteration. Due to its adequacy for unbalanced data and consolidation of Precision and Recall, the macro-averaged F1-score was selected as the validation metric. After their optimization, the models were retrained with the complete training sets and a final evaluation was performed with the evaluation sets.

**Support Vector Machine.** SVM [20] attempts to find a hyperplane that successfully segregates two classes in an $n$-dimensional space, where $n$ is the number of features. Even though it only inherently performs binary classification, a One-vs-All scheme was employed to handle multi-class classification. Table 2 summarizes the configuration.

Table 2. Summary of SVM configuration.

| Parameter | Value |
| --- | --- |
| Kernel | Linear |
| Loss Function | Squared Hinge |
| Dual | False |
| C | 0.001 to 0.1 |

The parameter search led to the use of the Linear kernel with the Squared Hinge loss function. Since the number of samples is significantly higher than the number of features across all datasets, Dual was set to *False* to solve the primal optimization problem.

This model relies on the C parameter, a value inversely proportional to the strength of the regularization. It was set to lower values on the larger datasets and higher values on the smaller datasets, in the range of 0.001 to 0.1.

**Extreme Gradient Boosting.** XGBoost [21] performs gradient boosting using an ensemble of decision trees. A level-wise growth strategy is employed to split nodes level by level, seeking to minimize a loss function. Table 3 summarizes the configuration.

Table 3. Summary of XGBoost configuration.

| Parameter | Value |
|---|---|
| Method | Histogram or Exact |
| Loss Function (Objective) | Cross-Entropy |
| Max Depth | 5 |
| Feature Subsample | 0.7 |
| Min Loss Reduction (Gamma) | 0.01 |
| Min Child Weight | 1.2 to 100.0 |
| Nº of Estimators | 60 to 80 |
| Learning Rate | 0.001 to 0.01 |

The acknowledged Cross-Entropy loss function was used for both binary and multiclass classification. To build the decision trees on the smaller datasets, the Exact method was utilized to account for all possible node splits. On the larger datasets, the Histogram method was selected because it computes fast histogram-based approximations.

The key parameters are the number of estimators and the learning rate. The first represents the number of decision trees, whereas the latter controls how quickly the model adapts its weights to the training data. Overall, the number of estimators was set to a relatively large value and the learning rate to a small value, avoiding a fast convergence to a suboptimal solution.

**Light Gradient Boosting Machine.** LightGBM [22] also utilizes an ensemble of decision trees to perform gradient boosting. A leaf-wise strategy is employed for a best-first approach, directly splitting the leaf with the maximum loss reduction. Consequently, despite having similar parameters to XGBoost, these have different effects on its performance. Table 4 summarizes the configuration.

Table 4. Summary of LightGBM configuration.

| Parameter | Value |
|---|---|
| Method | GOSS |
| Loss Function (Objective) | Cross-Entropy |
| Max Depth | 5 |
| Max Leaves | 25 |
| Feature Subsample | 0.7 |
| Min Loss Reduction (Split Gain) | 0.01 |
| L2 Regularization (Lambda) | 1.0 |
| Min Child Samples | 2 to 2000 |
| Nº of Estimators | 60 to 100 |
| Learning Rate | 0.001 to 0.04 |

The key advantage of this model is the ability to use Gradient-based One-Side Sampling (GOSS) to build the decision trees, which is computationally lighter than the remaining methods and therefore provides a faster and reliable convergence.

The Cross-Entropy loss function was also used and the learning rate was set to a small value on most datasets. However, the smaller and more unbalanced sets required it to be increased to counteract the shortage of training data.

### 3.5  Unsupervised Learning Models

Two unsupervised techniques were also selected because of their promising results in the surveyed work. Even though the developed models only perform one-class classification with unlabeled data, they can be compared to the remaining models in the binary scenario. Therefore, their optimization process was similar to the supervised approach, employing cross-validation to assess their configurations on unlabeled subsets.

**Isolation Forest.** An iForest [23] isolates anomalies through an ensemble of decision trees. The samples are repeatedly split by random values of random features until outliers are segregated from normal observations. Table 5 summarizes the configuration.

Table 5. Summary of iForest configuration.

| Parameter | Value |
|---|---|
| Nº of Estimators | 100 |
| Max Features | 1.0 |
| Max Samples | 100 to 250 |
| Contamination | 0.001 to 0.05 |

This model relies on the contamination ratio of the training set, which must not exceed 50%. Consequently, the number of samples intended to be anomalies must be lower than the number of remaining samples, otherwise outliers cannot be detected.

For 20-1, 21-1, 42-1 and 44-1, the ratio was set to the approximate percentage of malicious flows of the training sets. Even though 1-1-full and 34-1 do not fit the 50% requirement, 1-1-large, 1-1-medium and 1-1-small have exactly 50/50 proportions. Despite being theoretically suitable, the model underperformed with such high contamination. To overcome this obstacle, the samples with a malicious label were randomly subsampled to reduce the contamination of their training sets. The optimized ratio was 0.05, with approximately 5% malicious and 95% benign samples.

**Local Outlier Factor.** LOF [24] detects anomalies by measuring the local density deviation. This strategy identifies samples with a significantly lower density than their neighbors, which correspond to local outliers that would otherwise remain undetected.

Even though LOF only identifies anomalies on the initial data it receives by default, Novelty was set to *True* to enable it to detect outliers on new data, based on the neighborhoods computed in its training. Table 6 summarizes the configuration.

Table 6. Summary of LOF configuration.

| Parameter | Value |
| --- | --- |
| Novelty | True |
| Algorithm | K-Dimensional Tree |
| Metric | Euclidean |
| Leaf Size | 30 |
| Nº of Neighbors | 35 to 520 |
| Contamination | 0.001 to 0.05 |

The parameter search led to the values of the remaining parameters, as well as the use of the K-Dimensional Tree algorithm and the Euclidean metric. Regarding the contamination ratio of the training data, the approach employed for iForest was replicated.

The key parameter of this model is the number of neighbors, which regulates the size of the neighborhoods and therefore affects the measurement of the local density deviation. It was set to a higher value as the size of the dataset increased.

### 3.6  Deep Reinforcement Learning Model

To adapt the reinforcement learning methodology to the intrusion detection context, it was necessary to create a suitable training environment and develop a learning process for an agent. Due to the characteristics of this methodology, a manual optimization of several aspects was performed instead of cross-validation.

Regarding the training environment, when the agent observes a state and performs an action, predicting a class, it advances into the next state and provides a reward for the performed action. Due to the conclusions reached in [16], a simple 1/0 reward is calculated for correct/incorrect predictions.

Regarding the agent, an incremental episode-based learning process was developed, where each episode contains multiple steps (see Fig. 2). It was based on a DDQN [25] because this technique introduced several improvements to the training of an Artificial Neural Network (ANN). Therefore, the following concepts were applied:

- **Exploration** – During the initial training, the agent implements the Epsilon-Greedy method to choose predictions of the utilized ANN or random actions according to an exploration ratio. This method avoids a fast convergence to a suboptimal solution.
- **Experience Replay** – Instead of immediately updating the ANN's weights after an interaction with the environment, the agent stores those experiences in a finite memory. Then, a minibatch of past experiences is randomly sampled from the memory to train the ANN. Consequently, the interaction phase is logically separated from the learning phase, which mitigates the risk of catastrophic interference.
- **Target network** – Instead of using the same ANN for predicting the actions and the target values during experience replay, the agent employs two separate networks. An active network is continuously trained while a target network is used to calculate soft targets, being a copy of the first with delayed synchronization. This approach improves the generalization of the model by minimizing the instabilities inherent to the incremental training of an ANN.

In addition to the reward for the current action, a DDQN also calculates the expected future rewards during experience replay. However, since the correctness of future predictions is not relevant to the classification of a network flow, these were not calculated.

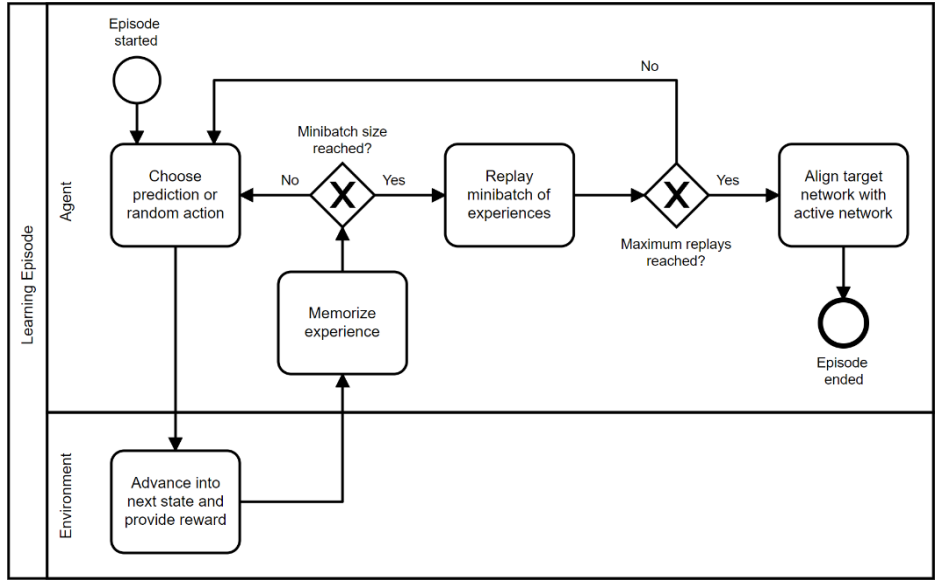

**Fig. 2.** Overview of DRL learning episode (Business Process Model and Notation).

Several parameters were manually optimized to regulate the developed learning process and the training of the agent. Table 7 summarizes the configuration.

The exploration rate was set to 0.2, which corresponds to 80% predictions and 20% random actions. The rate is decayed by 0.01 after each experience replay until a minimum of 0.05 is reached, effectively decreasing the random actions as the weights are adapted to the training data. The best balance between underfitting and overfitting was achieved with 2 replays per episode, each cycling through a minibatch for 20 epochs.

The size of a minibatch is the number of randomly sampled past experiences in an experience replay. Considering the small size of most utilized datasets, it was set to 2.5% of the size of a training set. Due to the greater size of the 1-1-full, 1-1-large and 1-1-medium datasets, this percentage was decreased. To strengthen the training of the agent, memory size was set to 1.5x the minibatch size, which corresponds to 3.75% of the size of a training set. Therefore, the agent can retrain with up to half of the already replayed experiences.

To perform the final evaluation, the most up-to-date network is retrieved from the agent. For that purpose, the learning process is stopped when the model's loss is stabilized. Stabilization is achieved when the average loss of the experience replays of the most recent episode is within the same range as the previous episodes. The number of previous episodes to compare and the stability range were set to 3 and 0.05, respectively, which gives margin for a slight variance.

Table 7. Summary of DRL learning process configuration.

| Parameter | Value |
|---|---|
| Exploration Rate | 0.2 |
| Exploration Rate Decay | 0.01 |
| Min Exploration Rate | 0.05 |
| Replays per Episode | 2 |
| Replay Epochs | 20 |
| Min Stable Episodes | 3 |
| Stability Range | 0.05 |
| Minibatch Size | 2.5% of set |
| Memory Size | 3.75% of set |

Regarding the active and target networks, both consist of a four-layered ANN. The Adam optimization algorithm is used to minimize the Cross-Entropy loss, with a learning rate of 0.001 to avoid a fast convergence to suboptimal weights.

The input layer node size is the number of utilized features, expressed as NF. Next, there are two hidden layers with 20 neurons each and the computationally efficient Rectified Linear Unit (ReLU) activation function. Finally, the binary output layer uses the Sigmoid activation function and a single node. For multi-class output, the layer is created with the Softmax function and a node size matching the total number of classes to be predicted, expressed as NC. Table 8 describes the employed structure.

Table 8. Employed ANN structure.

| Layer | Size | Activation |
|---|---|---|
| Dense | NF | - |
| Dense | 20 | ReLU |
| Dense | 20 | ReLU |
| Dense | 1 or NC | Sigmoid or Softmax |

## 4  Results and Discussion

This section presents and analyses the results obtained in the binary and multi-class classification scenarios, comparing the performance of the developed models.

### 4.1  Binary Classification

For the binary scenario, a comparison was performed between the F1-scores obtained in the cross-validation and the final evaluation. The obtained results are summarized in Tables 9 and 10, respectively.

In the 5-fold cross-validation, the supervised models, namely SVM, XGBoost and LightGBM, achieved scores near 100% when training with a large quantity of balanced data. The main distinction between the three models is visible on 21-1, where XGBoost only reached approximately 89.98%, despite SVM and LightGBM both obtaining 97.99%. On 34-1, a dataset unbalanced towards malicious flows, LightGBM obtained the highest score, a value of 99.73%.

In contrast with the supervised models, the scores of iForest and LOF were significantly lower on most of the larger datasets. Nonetheless, these unsupervised models achieved a good performance on the smaller and more unbalanced sets. LOF obtained better results than iForest on 1-1-small, 21-1 and 34-1. On 21-1, it surpassed XGBoost with a score of 91.66%. However, iForest outperformed LOF on all the remaining sets and even reached approximately 100% on 42-1.

Table 9. F1-scores of the binary cross-validation (5-fold average).

| Model | 1-1-full | 1-1-large | 1-1-medium | 1-1-small | 20-1 | 21-1 | 34-1 | 42-1 | 44-1 |
|---|---|---|---|---|---|---|---|---|---|
| SVM | **100** | **100** | **100** | **100** | **100** | **97.99** | 99.30 | **100** | **97.84** |
| XGBoost | 99.99 | 99.99 | 99.99 | 99.99 | **100** | 89.98 | 98.14 | **100** | **97.84** |
| LightGBM | **100** | 99.99 | 99.99 | 99.99 | **100** | **97.99** | **99.73** | **100** | **97.84** |
| iForest | 76.62 | 71.88 | 71.82 | 73.36 | 93.75 | 68.15 | 88.20 | **100** | 88.79 |
| LOF | 62.18 | 61.88 | 61.03 | 80.64 | 93.54 | 91.66 | 97.43 | 79.97 | 87.89 |

In the final evaluation, the supervised models achieved a good generalization. However, the lower scores on 20-1 and 21-1 indicate a slight overfitting on those smaller sets. On 44-1, the smallest of the analyzed datasets, only SVM increased its score.

A significant improvement is visible on the results of iForest on the larger sets, as well as on 20-1, where it reached 100%. This indicates it is well suited for the detection of anomalies on unseen data. On the other hand, LOF obtained lower scores on all datasets except 44-1. On this last set, its score was also increased to approximately 100%, possibly due to the small number of new samples to be classified.

The DRL model almost reached the results of the supervised models on the larger sets. However, it is pertinent to note that the smaller the training set, the lower the obtained score. This suggests that a large quantity of data is required for the developed learning process to be effective in an initial training.

Table 10. F1-scores of the binary evaluation.

| Model | 1-1-full | 1-1-large | 1-1-medium | 1-1-small | 20-1 | 21-1 | 34-1 | 42-1 | 44-1 |
|---|---|---|---|---|---|---|---|---|---|
| SVM | **100** | **100** | **100** | **100** | 95.43 | **94.42** | 99.43 | **100** | **100** |
| XGBoost | 99.99 | 99.99 | 99.99 | 99.99 | 95.43 | **94.42** | 98.84 | **100** | 96.28 |
| LightGBM | **100** | 99.99 | **100** | **100** | 95.43 | **94.42** | **99.76** | **100** | 96.28 |
| iForest | 96.46 | 94.80 | 94.68 | 95.37 | **100** | 89.95 | 75.08 | **100** | 90.91 |
| LOF | 53.46 | 53.40 | 54.66 | 80.18 | 89.95 | 87.45 | 96.80 | 49.96 | **100** |
| DRL | 99.91 | 99.91 | 99.97 | 99.98 | 78.49 | 83.28 | 98.65 | 83.31 | 75.39 |

Overall, the analyzed supervised and DRL models were reliable on most datasets, despite their slight performance decrease on some of the smaller sets. On the other hand, the unsupervised models were more advantageous for the smaller training sets, especially the ones highly unbalanced towards benign flows (see Fig. 3).

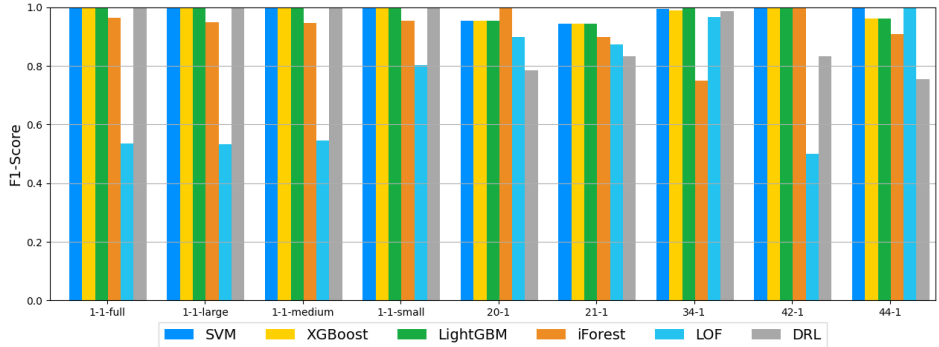

**Fig. 3.** Comparison of the F1-scores of the binary evaluation.

### 4.2  Multi-class Classification

For the multi-class scenario, an equivalent comparison was performed using the macro-averaged F1-scores, which are summarized in Tables 11 and 12. Due to the inability of unsupervised models to perform multi-class classification, these were not analyzed.

In the 5-fold cross-validation, the supervised models achieved very high scores on 34-1 and 44-1. LightGBM reached the highest score on 34-1, as in the previous scenario. On the other hand, very poor results were obtained on the particularly unbalanced sets. Since 1-1-full, 1-1-large and 42-1 contain minority classes with a very low number of samples, the models were not able to learn how to correctly classify them while training with 4/5 of the training sets.

**Table 11.** Macro-averaged F1-scores of the multi-class cross-validation (5-fold average).

| Model | 1-1-full | 1-1-large | 34-1 | 42-1 | 44-1 |
|---|---|---|---|---|---|
| SVM | **66.67** | **80.00** | 95.67 | 59.97 | **97.66** |
| XGBoost | 66.66 | **80.00** | 97.30 | 46.67 | 96.44 |
| LightGBM | 66.66 | **80.00** | **98.77** | **59.99** | **97.66** |

In the final evaluation, the supervised models reached scores of approximately 100% on 44-1 and similar results to the cross-validation on 34-1, which indicates a good generalization. However, their scores were decreased to near 66% on 1-1-large, due to the neglect of the underrepresented class. Furthermore, only LightGBM correctly classified one of the two minority classes of 42-1, whereas the remaining models failed to detect both. Since poor results were obtained on both validation and evaluation sets, the lack of training samples of those classes may be leading to underfitting.

The results obtained by the DRL model were very similar to the remaining models on most datasets, but significantly lower on 34-1 and 44-1. This indicates that the employed learning process cannot successfully account for multiple underrepresented classes during the initial training of the model.

Table 12. Macro-averaged F1-scores of the multi-class evaluation.

| Model | 1-1-full | 1-1-large | 34-1 | 42-1 | 44-1 |
|---|---|---|---|---|---|
| SVM | **66.67** | **66.67** | 95.89 | 33.31 | **100** |
| XGBoost | 66.66 | 66.66 | 95.59 | 33.33 | **100** |
| LightGBM | 66.66 | **66.67** | **99.64** | **66.65** | **100** |
| DRL | 66.64 | 66.64 | 63.75 | 33.38 | 88.38 |

Overall, the analyzed models achieved a good multi-class classification performance on the datasets with relatively balanced class proportions (see Fig. 4). The key obstacles remain the lack of training data and the underrepresented classes. Therefore, for these models to be able to distinguish between the different types of cyber-attacks, it is crucial to train them with a greater number of flows of each type.

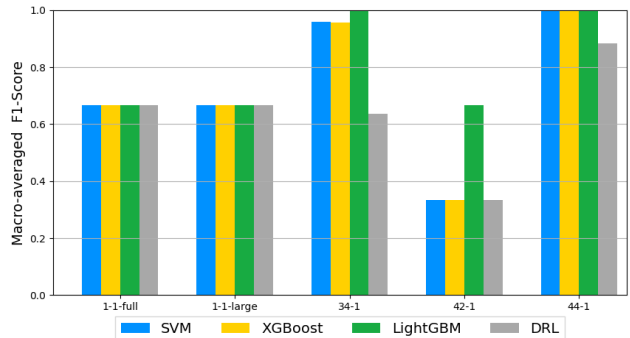

**Fig. 4.** Comparison of the macro-averaged F1-scores of the multi-class evaluation.

## 5    Conclusions

The developed work addressed IoT intrusion detection from a machine learning perspective. Nine malware captures of the IoT-23 dataset were utilized in a binary classification scenario and five of those in a multi-class scenario as well.

After a data preprocessing stage, three supervised models, SVM, XGBoost and LightGBM, two unsupervised models, iForest and LOF, and one DRL model based on a DDQN were analyzed and compared to assess their applicability to an IDS in an IoT system. Both a 5-fold cross-validation and a final evaluation were performed with the macro-averaged F1-score as the metric.

The supervised models achieved the most reliable performance in both scenarios, reaching higher scores when trained with a greater number of malware attack examples. LightGBM stood out for displaying the best generalization to several evaluation sets, especially in the multi-class scenario. Despite the significantly lower results of the unsupervised models, these seem advantageous for the detection of very low-frequency malware attacks. Furthermore, iForest achieved a good overall performance, which highlights its suitability for anomaly detection when trained with smaller and more unbalanced datasets.

The DRL model adapted to the intrusion detection context demonstrated that the reinforcement learning methodology can reach the performance of supervised techniques while also providing a learning process capable of continuously improving the detection. Therefore, the model can be adapted to changes in the traffic patterns, caused by updates to the network topology or by modifications to the cyber-attacks.

In the future, these three distinct types of machine learning techniques can be combined in an IDS to strengthen their benefits and overcome their individual drawbacks. Additionally, the development of DRL learning processes with rewards obtained from user feedback or other systems is a promising strategy to provide a more reliable and robust IoT intrusion detection.

**Acknowledgments.** The present work was done and funded in the scope of the European Union's Horizon 2020 research and innovation program, under project SeCoIIA (grant agreement no. 871967). This work has also received funding from UIDP/00760/2020.